\documentclass[3p,times]{elsarticle}

\usepackage{ecrc}

\volume{00}

\firstpage{1}

\journalname{Nuclear Physics A}

\runauth{T. Lappi}

\jid{nupha}

\usepackage{amssymb}
\usepackage[breaklinks]{hyperref}
\usepackage{amsmath}
\usepackage{mciteplus}

\newcommand{\qs}{Q_\mathrm{s}}

\newcommand{\as}{{\alpha_\mathrm{s}}}
\newcommand{\nc}{{N_\mathrm{c}}}
\newcommand{\ktt}{k_T} % scalar
\newcommand{\xt}{\mathbf{x}_T}
\newcommand{\yt}{\mathbf{y}_T}

\newcommand{\ut}{\mathbf{u}_T}
\newcommand{\vt}{\mathbf{v}_T}
\newcommand{\kt}{\mathbf{k}_T}

\newcommand{\nr}[1]{(\ref{#1})}
\newcommand{\ud}{\mathrm{d}}

\begin{document}

\begin{frontmatter}



% Your Title - please insert
\title{Multigluon correlations in JIMWLK}

% Principle author, and co-authors - please insert
\author{T. Lappi}

% Address - please insert
\address{Department of Physics
P.O. Box 35, 40014 University of Jyv\"askyl\"a, Finland and \\
Helsinki Institute of Physics,
P.O. Box 64, 00014 University of Helsinki, Finland
}

\dochead{}

\begin{abstract}
We discuss applications of the JIMWLK renormalization group 
equation to multigluon correlations in high energy collisions.
This includes recent progress in  computing the  energy dependence 
of higher point Wilson line correlators from the  JIMWLK 
renormalization group equation. We find that the large $\nc$
 approximation used so far in the phenomenological literature 
is not very accurate. On the other hand a Gaussian finite $\nc$
 approximation is surprisingly close to the full result.
We also discuss correlations at large rapidity separations, 
relevant for the ``ridge'' correlations observed in experiments.
\end{abstract}


\end{frontmatter}

% \linenumbers


\section{Introduction}\label{sec:intro}
The physics of high energy hadronic or nuclear collisions is dominated
by the gluonic degrees of freedom of the colliding particles. 
These
small~$x$ gluons form a dense nonlinear system that is, at high enough
$\sqrt{s}$, best described as a classical color field and quantum 
fluctuations around it. 
The color glass condensate (CGC,
for  reviews see \cite{Gelis:2010nm,*Lappi:2010ek})
is an effective theory developed around this idea. It gives an 
universal description of the small~$x$ degrees of freedom 
that can equally well be applied to small~$x$
DIS as to  dilute-dense (pA or forward AA) and dense-dense (AA or very 
high energy pp) hadronic collisions. 
The nonlinear interactions of the small $x$ gluons generate dynamically 
a new transverse momentum scale, the saturation scale $\qs$,
that grows with energy.
At high enough energy the color glass condensate is thus a one-scale 
system, characterized by a dominant momentum scale $\qs$ that is hard
 enough to 
justify a weak coupling calculation. The scale $\qs$ dominates both the
gluon spectrum and multigluon correlations. The nature
of a unique saturation scale as both the typical gluon transverse 
momentum and as the correlation length $1/\qs$ differentiates the CGC
qualitatively from the high-$x$ part of the wavefunction.

The most convenient parametrization of the dominant gauge field
is in terms of Wilson lines that describe the eikonal propagation 
of a projectile through it. 
The Wilson lines are drawn from a probability distribution, whose
dependence on rapidity is described by the JIMWLK renormalization
group equation.
It reduces, in a large
$\nc$ and mean field approximation, to the 
BK~\cite{Balitsky:1995ub,*Kovchegov:1999yj} equation and further, in the
dilute linear regime, to the BFKL one.

\section{Correlations in a dilute-dense collision}\label{sec:multipt}

\begin{figure}
\includegraphics[width=0.5\textwidth]{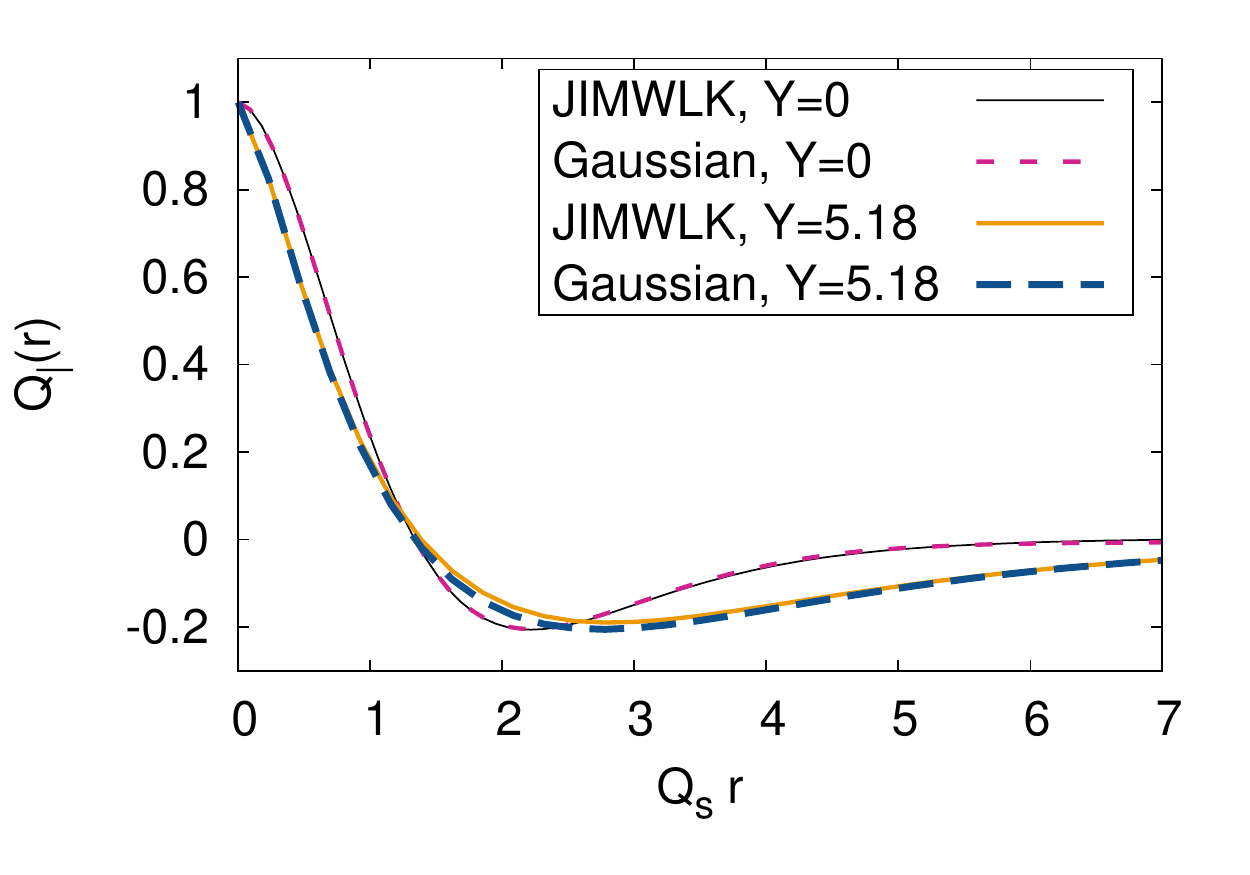}
\includegraphics[width=0.5\textwidth]{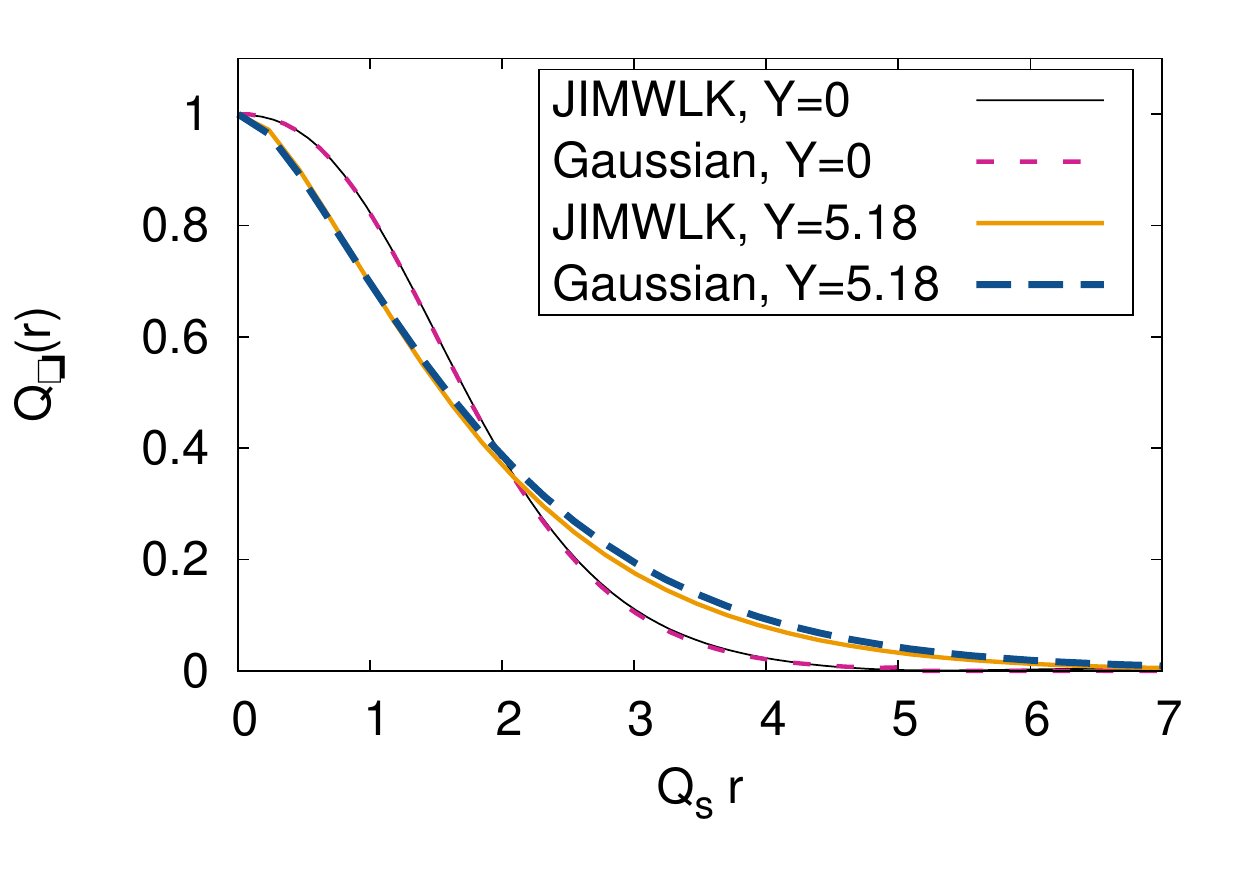}
\caption{\label{fig:4pt}
The JIMWLK result for the quadrupole correlator compared to the
Gaussian approximation. Shown are the initial condition (MV model) at 
$y=0$ and the result after $5.18$ units of evolution 
in rapidity, for  the ``line'' (left) and
``square'' (right) coordinate configurations.
Figures from Ref.~\cite{Dumitru:2011vk}.
}
\end{figure}

One of the more striking signals of  saturation physics
at RHIC is 
seen in the relative azimuthal angle $\Delta \varphi$ dependence 
of the dihadron correlation function, where the 
$\Delta \varphi\approx \pi$ back-to-back peak is seen to 
be suppressed in dAu-collisions compared to pp collisions at
the same kinematics~\cite{Adare:2011sc,*Braidot:2011zj}.
The CGC description of this correlation starts from a large~$x$
parton pair propagating eikonally through the target. 
For a dilute target momentum conservation causes a peak in the
correlation function at $\Delta \varphi \approx \pi$, 
whereas for a dense one the intrinsic transverse momentum, of the order of 
$\qs$, causes the peak to disappear for semihard momenta.
To calculate the matrix element for this process 
one needs target expectation values of  products of Wilson line
operators, such as the dipole and the quadrupole
\begin{equation}\label{eq:dipquad}
\hat{D}(\xt,\yt) = 
\frac{1}{\nc} \mathrm{Tr} U(\xt) U^\dagger(\yt)
\quad 
\hat{Q}(\xt,\yt,\ut,\vt) = 
\frac{1}{\nc} \mathrm{Tr} U(\xt) U^\dagger(\yt)
U(\ut) U^\dagger(\vt).
 \end{equation}
For practical phenomenological work it would be extremely convenient to be 
able to express these higher point correlators in terms of the dipole,
which is straightforward to obtain from the BK equation.
In the phenomenological literature so 
far~\cite{Tuchin:2009ve,*Albacete:2010pg} this has been done using
a ``naive large $\nc$'' approximation where the quadrupole is assumed 
to be simply a product of two dipoles.
A more elaborate scheme would be a ``Gaussian'' approximation
(``Gaussian truncation'' in~\cite{Kuokkanen:2011je}), where
one assumes the relation between the higher point functions and the
dipole to be the same as in the (Gaussian) MV model. The 
expectation value of the quadrupole operator
in the MV model has been derived e.g. in Ref.~\cite{Dominguez:2011wm}.

In Ref.~\cite{Dumitru:2011vk} the validity of these approximations 
was studied by comparing them numerically to the 
solution of the JIMWLK equation.
As studying the full 8-dimensional phase space for the quadrupole operator
would be cumbersome, we have concentrated on two
special coordinate configurations. The 
``line'' configuration is defined by taking  $\ut = \xt$ and $\vt = \yt$,
with $r= |\xt-\yt|$ and the ``square'' by taking $\xt,\yt,\ut,\vt$ as the
corners of a square with side $r$.

\begin{figure}
\includegraphics[width=0.5\textwidth]{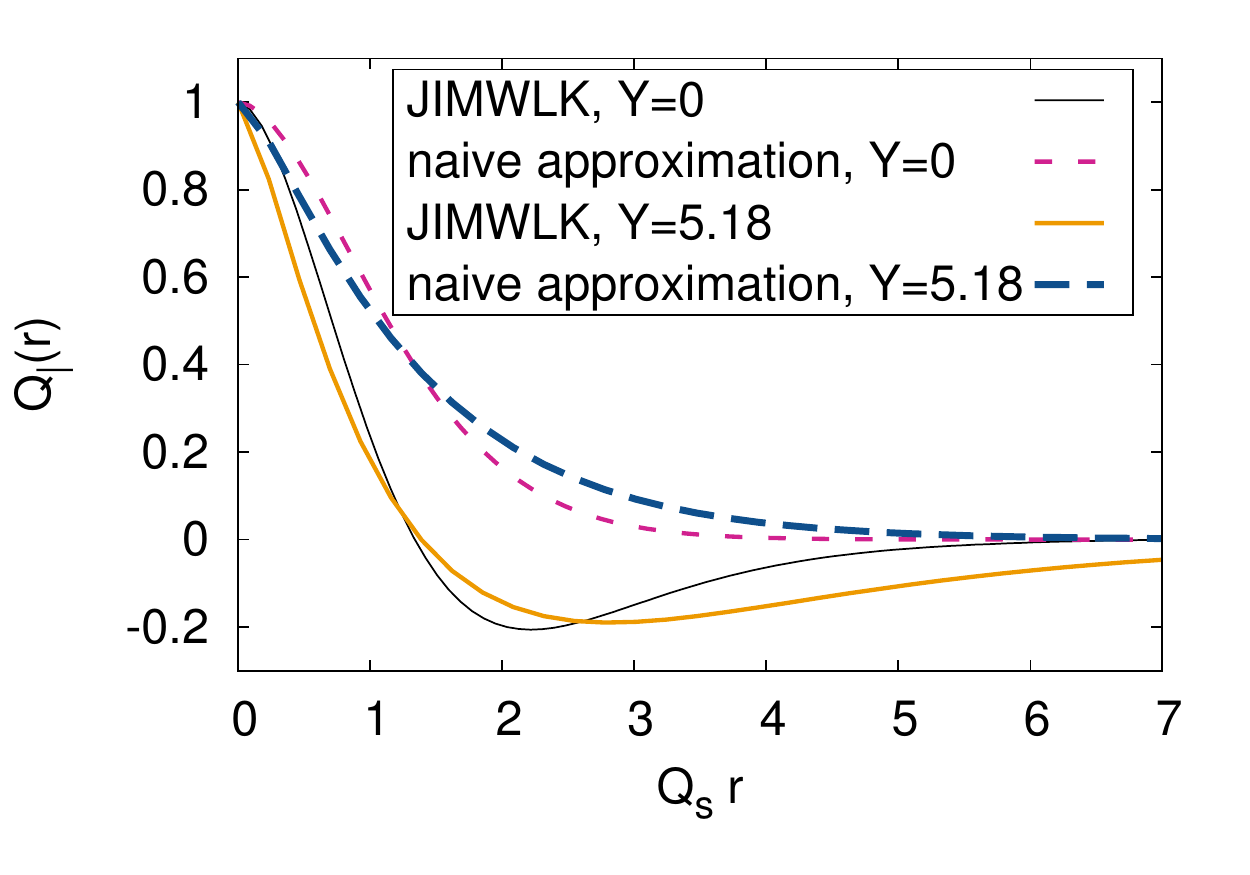}
\includegraphics[width=0.5\textwidth]{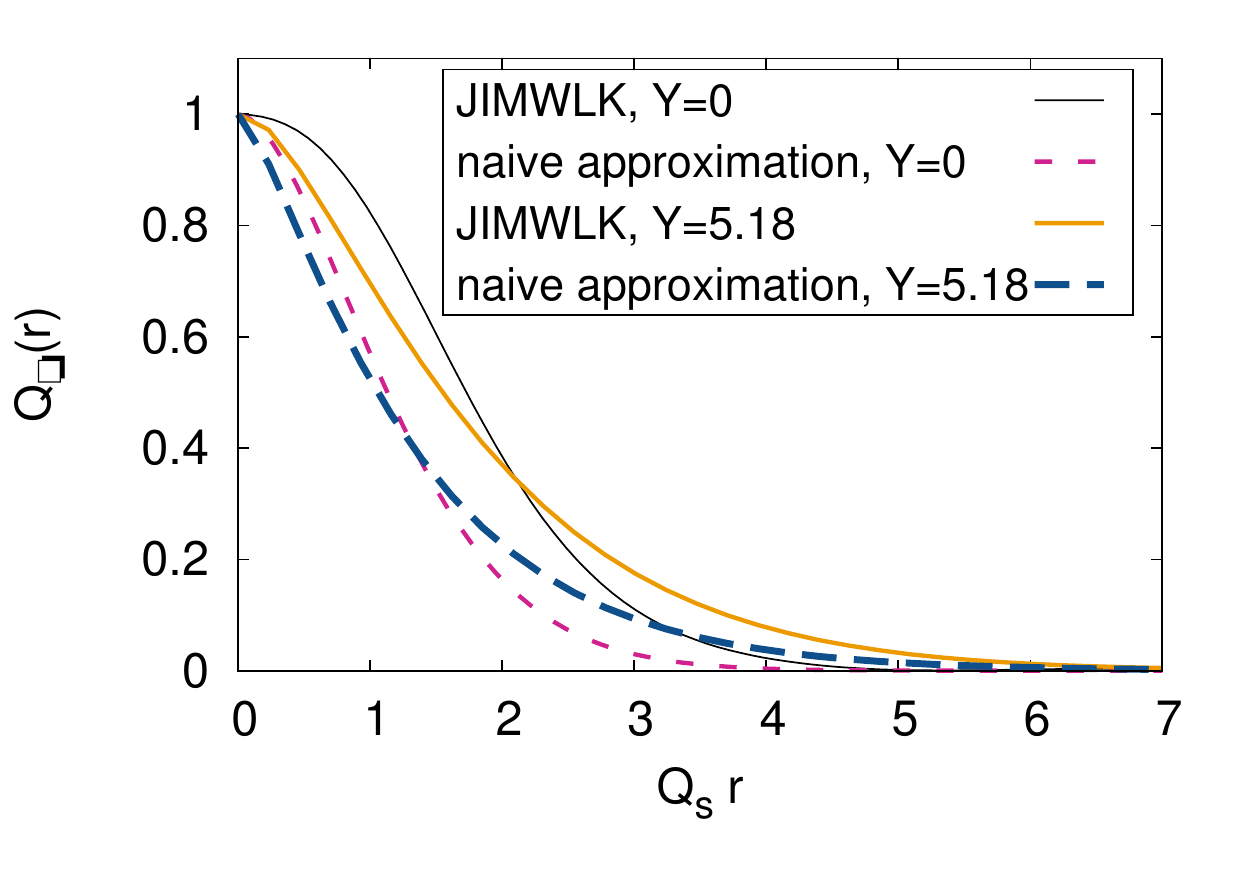}
\caption{\label{fig:4ptn}
The JIMWLK result for the quadrupole correlator compared to the
``naive large $\nc$'' approximation. Shown are the initial condition (MV model) at 
$y=0$ and the result after $5.18$ units of evolution 
in rapidity, for the ``line'' (left) and
``square'' (right) coordinate configurations.
Figures from Ref.~\cite{Dumitru:2011vk}.
}
\end{figure}

Our most important results in Ref.~\cite{Dumitru:2011vk} 
for the quandrupole expectation 
value are shown in Figs.~\ref{fig:4pt} and~\ref{fig:4ptn},
with a comparison of the initial and evolved (for 5.18 units in $y$)
results to the approximations.
The MV-model initial condition $y=0$ satisfies the Gaussian approximation 
by construction. Figure~\ref{fig:4pt} shows that the Gaussian approximation
is still surprisingly well conserved by the evolution. A possible explanation
for this based on  the structure of the JIMWLK equation has recently 
been proposed in~\cite{Iancu:2011ns,*Iancu:2011nj}. The naive large 
$\nc$ approximation,
on the other hand,  fails already at the initial condition, as shown 
in Fig.~\ref{fig:4ptn}. This stresses
the importance of the various SU(3) group structure constraints 
violated in this approach. Crucially for
the phenomenological consequences,
even the characteristic length/momentum scale differs by factor $\sim 2$
from the actual result.

This result does not yet fully address the effect on the
measurable dihadron cross section.
For that one must convolute the Wilson line operators
with the $q\to qg$ splitting wavefunction. This  nontrivial
numerical task is still work in progress,
discussed in this conference in \cite{HMHO12}.

\section{Unequal rapidity correlations and the ridge}\label{sec:neqy}

Calculating multigluon correlations in a collision of 
two dense gluonic systems requires including the JIMWLK
evolution for both of the colliding projectiles
(for a more formal discussion see \cite{Gelis:2008sz,*Lappi:2009fq}).
Of particular phenomenological interest are correlations
at large rapidity separations, because they are directly responsible
for the ``ridge'' long range rapidity correlations observed 
in nucleus-nucleus and high multiplicity proton-proton collisions.
In spite of this, most practical work on the JIMWLK equation
has so far concentrated on correlations of Wilson lines at one
rapidity. 

\begin{figure}
\begin{center}
\resizebox{0.99\textwidth}{!}{
\includegraphics[width=5cm,clip=true]{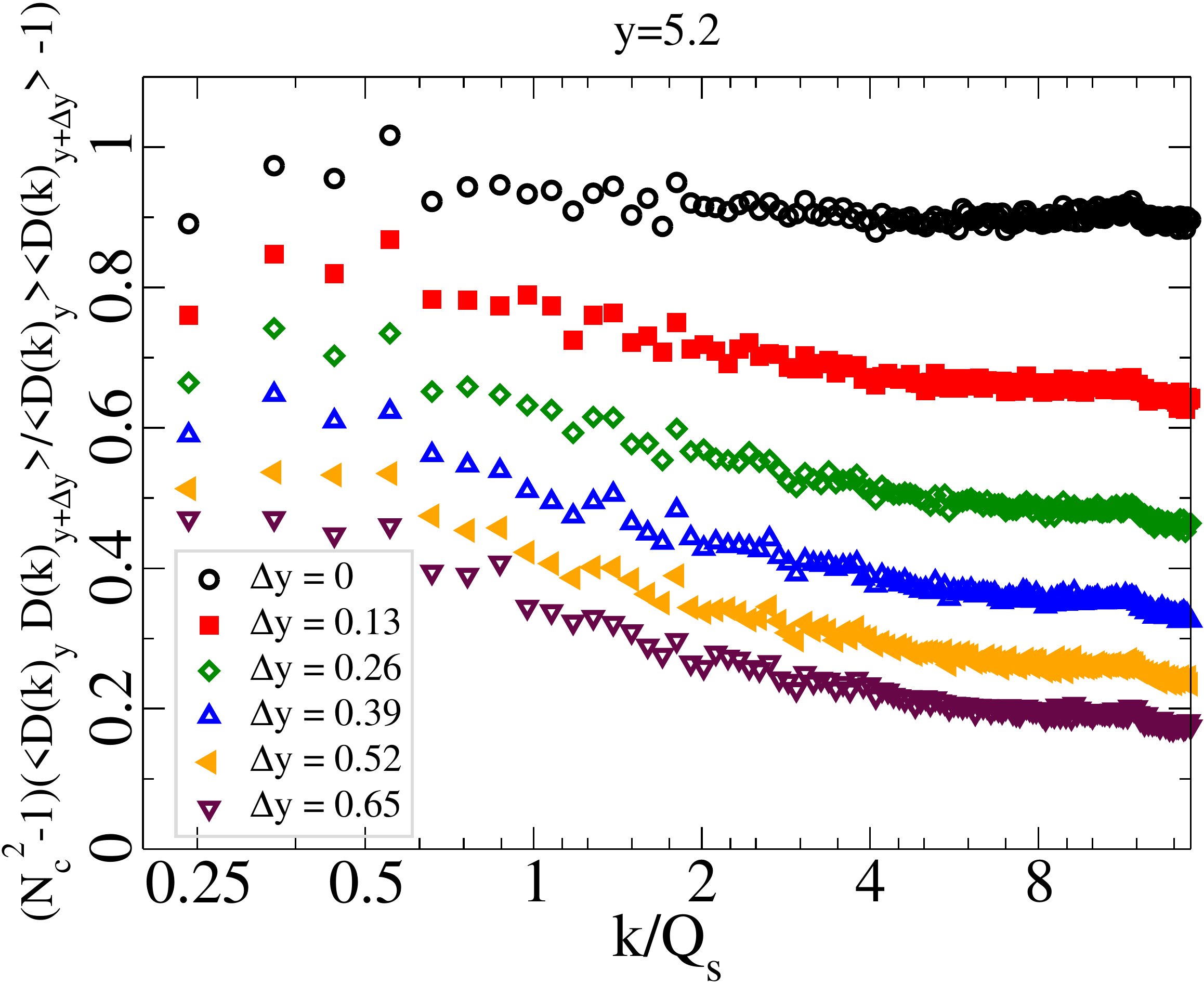}
\rule{1cm}{0pt}
\includegraphics[width=5cm,clip=true]{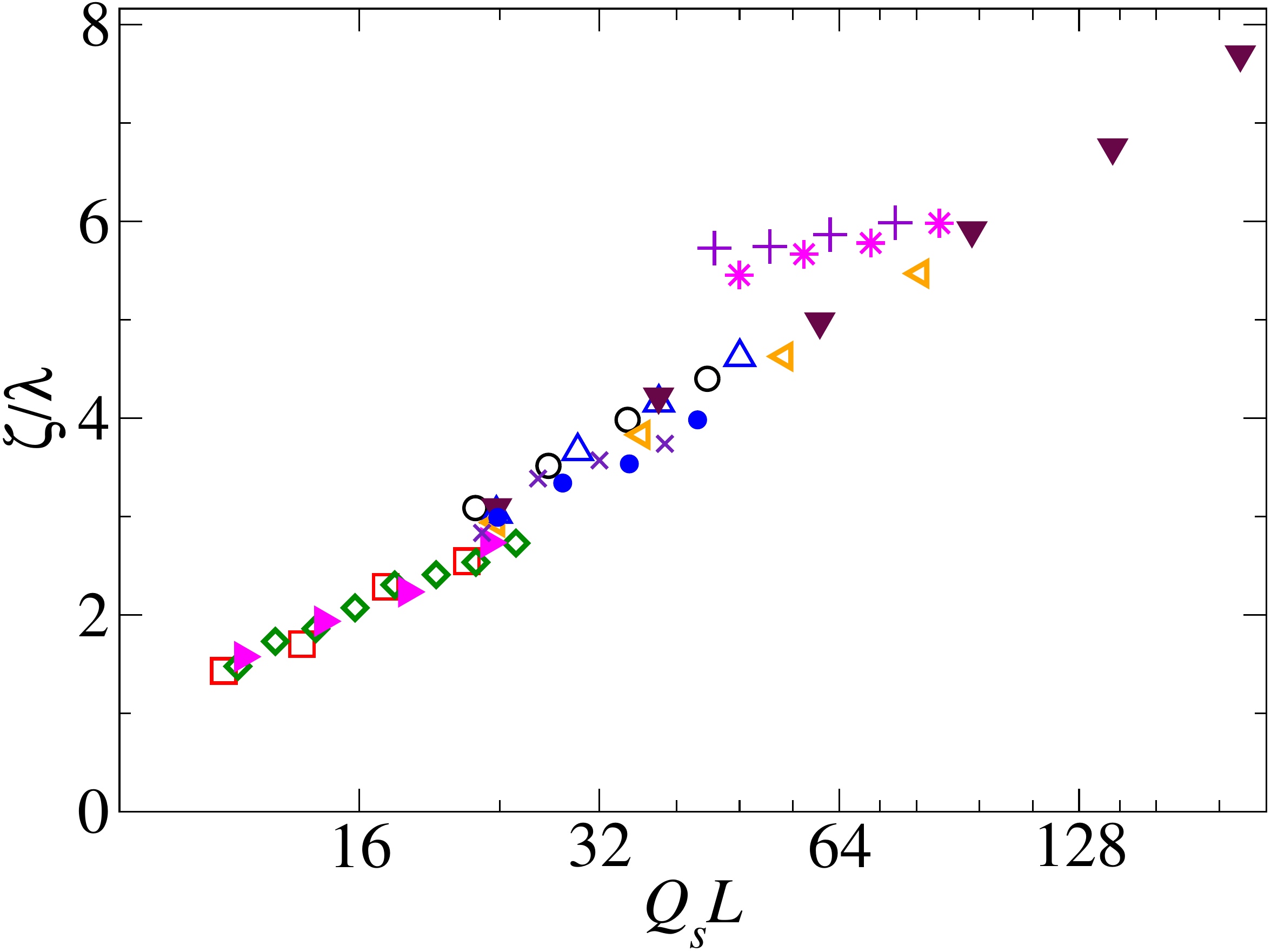}
}
\end{center}
\caption{
Left:  
Unequal rapidity correlation \nr{eq:neqycorr} as a function of
$\ktt/\qs$ for different rapidity separations.
Right:
Dependence of the decorrelation speed $\zeta$ on $\qs L$, where 
$L$ is the linear size of the system.
} \label{fig:neqy}
\end{figure}

The existing applications  of the CGC framework to calculations
of the ridge~\cite{Dumitru:2008wn,*Dusling:2009ni,Lappi:2009xa,%
*Dumitru:2010iy,*Dusling:2012ig} 
have used a simplified argument based on the MV model, where the
change in the effective color charge in one infinitesimal step
in rapidity is independent of the color charge. This leads
to an unequal rapidity correlation function of Wilson line operators
which is independent of the rapidity separation between the 
produced particles. Consider for
example the following correlation function, which is similar to the
one appearing in the calculation of the ridge correlation in the
$\ktt$-factorized approximation \cite{Dumitru:2008wn,*Dusling:2009ni}:
\begin{equation}\label{eq:neqycorr}
(\nc^2-1)
\left[
\frac{
	\left\langle 
	\hat{D}(\kt)_{y} 
	\hat{D}(\kt)_{y+\Delta y} 
	\right\rangle
}{	 
	\left\langle 
	\hat{D}(\kt)_{y} 
	\right\rangle
	\left\langle 
	\hat{D}(\kt)_{y+ \Delta y} 
	\right\rangle
}
-
 1
\right],
\end{equation}
where $\hat{D}(\kt)$ is the Fourier-transform of the dipole 
operator\nr{eq:dipquad}. In the  MV model 
calculation of \cite{Dumitru:2008wn,*Dusling:2009ni} this quantity would be
one independently of $\Delta y$.
This naturally leads to a very long range correlation in rapidity
between produced gluons, as seen in~\cite{Dumitru:2008wn,*Dusling:2009ni,Lappi:2009xa,%
*Dumitru:2010iy,*Dusling:2012ig}. Genuine JIMWLK evolution should be
expected to decorrelate the Wilson line operators, at a characteristic
rapidity scale $\Delta y \sim 1/\as$. In 
Fig.~\ref{fig:neqy} we show preliminary results for 
the correlation function \nr{eq:neqycorr}. On the left is plotted the
correlation function itself at different rapidity separations
$\Delta y$. There is only a mild dependence on the transverse momentum
$\ktt$, as expected from the  MV model calculation. The decorrelation seems, 
however, to be relatively fast. This is further quantified
on the right, where we show the result for the 
decorrelation speed $\zeta$ in  an exponential decay fit
\begin{equation}
\frac{
	\int \ud^2\kt \kt^4
	\left[
	\left\langle 
	\hat{D}(\kt)_{y} 
	\hat{D}(\kt)_{y+\Delta y} 
	\right\rangle
	- 
	\left\langle 
	\hat{D}(\kt)_{y} 
	\right\rangle
	\left\langle 
	\hat{D}(\kt)_{y+ \Delta y} 
	\right\rangle
	\right]
}{
	\int \ud^2\kt \kt^4
	\hat{D}(\kt)_{y}
}
\sim 
\exp \{ -\zeta y \}.
\end{equation}
Here the weighting with $\ktt^2$ is chosen to match the unintegrated
gluon distribution $\sim \ktt^2 D(\ktt)$
appearing in a
$\ktt$-factorized calculation 
of gluon production. In Fig.~\ref{fig:neqy} the decorrelation 
speed $\zeta$, explicitly $\sim \as$, is compared to the natural 
evolution speed $\lambda$, defined by $\lambda \equiv \ud \ln \qs^2
/\ud y$, measured 
in the same JIMWLK simulation. The decorrelation is seen to be surprisingly
fast. What is more worrying, there seems to be a slight logarithmic 
dependence on the infrared cutoff given by the system size $L$, which 
persists over a variety of different initial conditions, lattice sizes,
values at which the running coupling is frozen in the infrared 
and other parameters varied in the calculations. At the moment there
appears to be no natural interpretation of this dependence,
leaving the implications for ridge phenomenology uncertain.

\section{Conclusion}\label{sec:conc}

We have here argued that multiparticle correlations provide unprecedented
experimental insight into the details of nonlinear QCD dynamics 
at small $x$. In particular, they make it necessary to go beyond the
mean-field BK equation and use the full JIMWLK equation.
Azimuthal angle correlations of particles produced in a collision
with a dilute probe and dense, saturated target are sensitive to 
multipoint functions of Wilson lines in the target wavefunction.
We have shown that a ``naive large $\nc$'' approximation used in the
literature, where the quadrupole operator is assumed to be
a simple product of dipoles, is far from the true finite $\nc$ result.
A Gaussian approximation, based on the MV model, turns out to be 
surprisingly close to the result from JIMWLK evolution.
We have also discussed preliminary results on unequal rapidity 
correlations which are needed for a proper CGC calculation
of the ``ridge'' correlation in high energy collisions. We have
shown that at least some observables display a problematic
infrared behavior of the decorrelation speed in rapidity.

\paragraph{Acknowledgements} This work  has been supported by the Academy of Finland, 
projects 141555 and 133005, and by computing resources from
CSC -- IT Center for Science in Espoo, Finland.

\bibliographystyle{h-physrev4mod2M}
\bibliography{spires}

\end{document}